\documentclass{aastex631}
\usepackage{amsmath}

\newcommand\Ic{I_\mathrm{c}}

\begin{document}

\title{Non-zero phase-shifts of acoustic waves in the lower solar atmosphere measured from realistic simulations and their role in local helioseismology\footnote{Current Version: 0.00}}

\author[0000-0003-2678-626X]{M. Waidele}
\affiliation{W. W. Hansen Experimental Physics Laboratory, Stanford University, Stanford, CA 94305-4085, USA}

\author[0000-0002-6308-872X]{Junwei Zhao}
\affiliation{W. W. Hansen Experimental Physics Laboratory, Stanford University, Stanford, CA 94305-4085, USA}

\author[0000-0003-4144-2270]{I. N. Kitiashvili}
\affiliation{NASA Ames Research Center, Moffett Field, Mountain View, CA 94035, USA}

\begin{abstract}

Previous studies analyzing the evanescent nature of acoustic waves in the lower solar atmosphere, up to 300\,km above the photosphere, have shown an unexpected phase shift of an order of 1\,s between different heights. Those studies investigated the spectral line \ion{Fe}{1} 6173.3\,\AA, commonly used for helioseismic measurements. Such phase-shifts can contribute to a misinterpretation of the measured travel times in local helioseismology, complicating inferences of, e.g., the deep meridional flow. In this study, we carry out phase-shift computations using a simulated, fully radiative, and convective atmosphere from which the \ion{Fe}{1} 6173.3\,\AA\ line is synthesized. The resulting phase-shifts as functions of frequency across multiple heights show non-zero values in evanescent waves, similar to what was found in observational data. Comparing the Doppler-velocities estimated from the synthesized absorption line with the true velocities directly obtained from the simulated plasma motions, we find substantial differences in phase-shifts between the two. This leads us to hypothesize that the non-adiabaticity of the solar atmosphere yields extra phase-shift contributions to Doppler velocities. Finally, computing phase-differences for different viewing angles reveals a systematic center-to-limb variation, similar to what is present in observations. Overall, this study helps to improve our understanding of the physical cause of the helioseismic center-to-limb effect.

\end{abstract}

\section{Introduction} \label{sec:introduction}
Meridional circulation is a key component in explaining the solar cycle mechanism \citep[e.g.,][]{1961ApJ...133..572B, 2015LRSP...12....4H}. Although surface measurements of the meridional circulation were reliably measured by various methods \citep[e.g.,][]{1993SoPh..147..207K, Hathaway1996,Roudier2018}, the estimates of deep circulation still carry large uncertainties to this day, presenting an unclear and conflicting picture among different authors \citep[e.g.,][]{2013ApJ...774L..29Z, 2013ApJ...778L..38S, 2015ApJ...813..114R, 2015ApJ...805..133J, 2017ApJ...849..144C, 2017PhDT.......153B, 2020Sci...368.1469G}. Inferences of the deep meridional circulation rely on helioseismic measurements that can be inverted for depth-dependent velocity profiles \citep[see, e.g.,][]{2002RvMP...74.1073C, 2005LRSP....2....6G}. Aside from complications with the complexity of inversion techniques, especially in active regions \citep{2012SoPh..279..323K}, the helioseismic measurements themselves carry uncertainties and systematic effects. 
For instance, \citet{2003ESASP.517..259D} found a significant decrease with latitude in travel times of waves with same travel distances. When looking for asymmetries between wave travel times in opposite traveling directions (i.e. travel-time differences), \citet{2009ASPC..416..103D} showed that a systematic effect affects measurements performed along the equator. \citet{2012ApJ...749L...5Z} expanded this procedure to the entire solar disk and highlighted the presence of systematic center-to-limb variations in the helioseismic travel times. Although these results allow for an empiric treatment of the effect, its underlying cause is not well understood. An important component of the center-to-limb effect is its variation with the conceived atmospheric height. \citet{2016SoPh..291..731Z} investigated the effect due to geometric foreshortening, which may introduce perturbations to wave travel times due to a systematic shortening of travel distances as measurements are taken increasingly closer to the solar limb, and found that mean travel times are weakly affected and travel time differences remained largely unchanged. 
Another important characteristic of the center-to-limb effect was found by \citet{2018ApJ...853..161C}, who demonstrated that the effect strongly depends on the frequency of acoustic waves, where the maximum perturbation to the travel-time differences  forms around $3.7\,$mHz.

Recently, \citet{2022ApJ...933..109Z} reported an important finding that acoustic waves in the lower solar atmosphere carry unexpected phase-shifts that can result in spurious travel-time shifts of up to $1\,$s, which may help understand the nature of the helioseismic center-to-limb effect. Such systematic phase-shifts must be taken into account for accurate inferences of the deep meridional flow. Other helioseismic results such as supergranular and sunspot flows may be affected as well. 
In this context, evanescent waves are understood as standing waves with frequencies below the acoustic cut-off frequency, which exhibit an exponentially decaying amplitude as they reach into higher atmospheric layers.
Thoroughly explaining the extra phase-shifts carried by evanescent acoustic waves is non-trivial. \citet{2022ApJ...933..109Z} postulated the following:
An evanescent acoustic wave present in the lower solar atmosphere causes a periodic compression of local plasma volume. The resulting temperature oscillation is likely out of phase in the non-adiabatic atmosphere near and above the Sun's photosphere, which can cause a phase lag between temperature perturbations and local plasma compressions. Thus, the Doppler velocity derived from the measured intensity likely carries a spurious phase-shift relative to the plasma velocity. 

To investigate the origin of the extra phase shift in helioseismic travel times, we perform an analysis of 3D radiative simulations and  the corresponding synthesized \ion{Fe}{1} 6173\,\AA\ line profiles that show a systematic increase of the line formation height for areas closer to the solar limb \citep{Kitiashvili2015}. In this paper, we apply similar analysis procedure as described by \citet{2022ApJ...933..109Z} to investigate the evanescent nature of acoustic waves using the same simulation as \citet{Kitiashvili2015}. This paper is structured as follows: Section \ref{sec:dataandsimulation} describes observational data and simulations, and in Section \ref{sec:methods} we explain the methods used leading up to the computation of instantaneous wave phases. Finally, we present our results in Section \ref{sec:results}, discuss them in Section \ref{sec:discussion}, and draw our conclusion in Section \ref{sec:conclusion}.

\section{Data and simulation} \label{sec:dataandsimulation}
In this study, we evaluate the instantaneous wave phases in the form of phase-differences $\delta\phi$, measured from a set of realistic numerical simulation data of the solar atmosphere. The simulation data, obtained using the \textit{StellarBox} code developed at NASA/Ames Research Center \citep{Wray2015,Wray2018}, were previously used to study center-to-limb effect \citep{Kitiashvili2015}. This dataset is also used to simulate the line profiles of \ion{Fe}{1} 6173\AA, the same line used by the Helioseismic and Magnetic Imager (HMI) onboard the Solar Dynamics Observatory (SDO) \citep{2012SoPh..275..229S}, through employing the \textit{SPINOR/STOPRO} code \citep{1992A&A...263..312S, 2012A&A...548A...5V}. For the purpose of validating the phase-differences measured from the simulation dataset, we also compute $\delta\phi$ using observations by the Interferometric BIdimensional Spectropolarimeter  \citep[IBIS;][]{2006SoPh..236..415C}. The same IBIS dataset was well described by \citet{2010ApJ...721L..86R} and \citet{2022ApJ...933..109Z}, so we only present a quick summary here in section \ref{sec:ibisdata}. 

\subsection{IBIS data} \label{sec:ibisdata}
The IBIS observations used in this study comprise a spectrally resolved and fully tracked patch of quiet-Sun around a sunspot. The observations were made on 2007 June 8 using IBIS installed at the Dunn Solar Telescope of the National Solar Observatory, Sacramento Peak, New Mexico, USA \citep{2006SoPh..236..415C}. IBIS features a spectral resolution of $25\,$m\AA, used to resolve the \ion{Fe}{1} $6173.3\,$\AA, a line commonly observed for helioseismic studies. The patch is tracked for a duration of $7\,$hours, with a cadence of $47.5\,$s, a spatial resolution of $0\farcs330$, and a field of view of $60\times60\,$Mm. Line-of-sight (LOS) Doppler velocities of local plasma motions are obtained by measuring the shift of the spectral line. We refer the readers to the detailed description of the Doppler shift calculations to \citet{2007ApJ...654L.175R, 2010ApJ...721L..86R}. Since the line is spectrally resolved, we can extract velocities within the line wings at different intensity levels that can roughly be translated to atmospheric heights. 

\subsection{Simulation data} \label{sec:simulationdata}
In this study we use a three-dimensional hydrodynamic radiative simulations obtained with the \textit{StellarBox} code \citep{Wray2015,Wray2018}. The computational model uses a large-eddy simulation (LES) treatment of subgrid turbulent transport. The radiative transfer has been computed in the local thermodynamic equilibrium (LTE) approximation. A three-dimensional multi-spectral-bin long-characteristics method is used to calculate the radiative transfer between fluid elements. Thereby, the ray-tracing transport is performed using the \citet{1964CR....258.3189F} method with 14 rays after ionization and excitation of all abundant species are taken into account. 
The boundary conditions are periodic in lateral directions, while the top boundary is open to mass, momentum, and energy and radiation fluxes. The bottom boundary is open only for radiation and has the energy input from the interior of the Sun. We used a standard solar model of the interior structure and the lower atmosphere \citep{1996Sci...272.1286C} and randomly distributed velocity perturbations with amplitude of few cm/s. 
Once the modeling reaches a steady state condition, the simulation exhibits observed dynamics and turbulent properties of a quiet-Sun region \citep[e.g.,][]{Kitiashvili2011,Kitiashvili2012,Kitiashvili2023,Hathaway2015}.

\begin{figure}[h!]
    \centering
    \includegraphics[width=7.in]{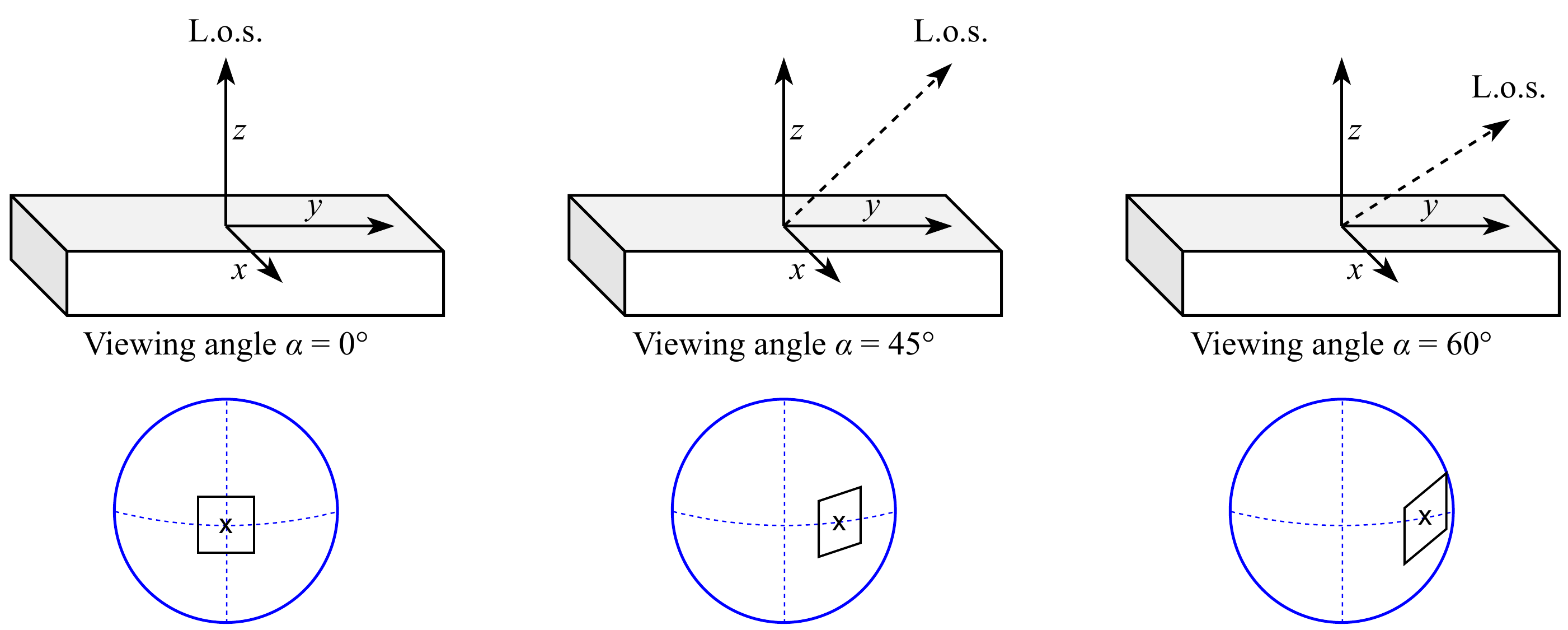}
    \caption{Illustration of the viewing angle geometry. The upper row shows the simulation boxes with coordinate frames on the top. In the bottom row, the respective theoretical disk location is illustrated. From left to right: $\alpha = 0^\circ, 45^\circ, 60^\circ$. For $\alpha=0^\circ$, the LOS direction coincides with the $z$-axis, while for the larger $\alpha$ the LOS is situated in the $z$--$y$--plane.}
    \label{fig:viewingangledrawing}
\end{figure}

The computational domain covers $5.2$\,Mm of upper layer of the convection zone and an atmospheric height of $1\,$Mm relative to the effective temperature of the Sun from 1D solar model S \citep{1996Sci...272.1286C}. Since we are only interested in waves above the photosphere, we limit ourselves to a height of $400\,$km with a vertical grid spacing of $12.1\,$km. In this work, we use simulation data for $1.05\,$hours with a temporal cadence of $50\,$s. To mimic the effect of different locations of the simulated area on the solar disk, we also use synthetic data sets obtained with a viewing angle $\alpha = 0^\circ$ (disc center), $45^\circ$, and $60^\circ$ (Figure \ref{fig:viewingangledrawing}), in which original models of the solar atmospheres have been projected to the corresponding line of sight (LOS).

\section{Methods} \label{sec:methods}
Solar oscillations are observed using Doppler velocities, calculated from the Doppler shift of an atomic absorption line relative to a reference profile. Different methods have been developed to estimate the Doppler shift, and ultimately, the line profile has to be measured as a function of wavelength. In this study, we consider two different velocity-deriving procedures, the bisector method and the HMI-like algorithm (see Sec.~\ref{sec:estimatingvelocities} for details). 
In addition, we utilize available 3D simulations to compare the simulated LOS flow field (`true' LOS velocities, $v_{\rm true}^{\rm LOS}$) and the  Doppler shift derived from synthetic iron line profiles to investigate discrepancies between them. For an accurate comparison of LOS velocities obtained directly from the numerical model and derived from synthesized spectroscopic data, we have to define the relative height for both quantities.
The proposed method allows us to approximate the intensity-height relation, translating intensity from the line wings $N = I(\lambda)/\Ic$ to atmospheric height $z$ ($\Ic$ is the continuum intensity). Afterwards, we compute the power spectra for both LOS velocities, which is necessary for the filter that is applied to consider only acoustic modes. 

\subsection{Estimating velocities} \label{sec:estimatingvelocities}

\subsubsection{Bisector}
\begin{figure}[h!]
    \centering
    \includegraphics[width=7.1in]{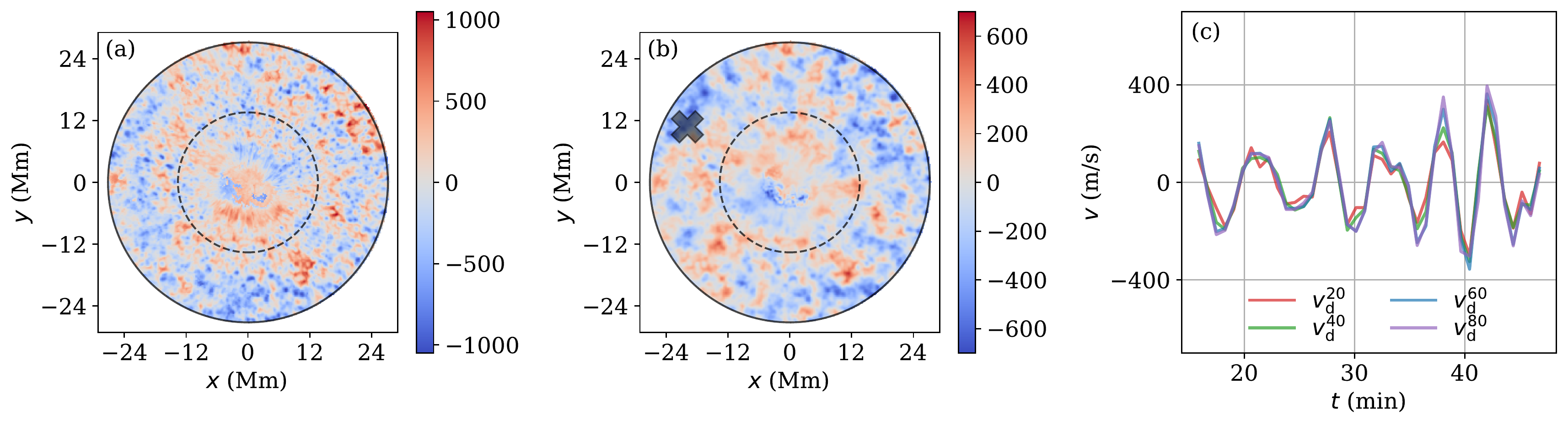}
    \caption{Doppler velocities extracted from the IBIS dataset. Panel (a) shows a snapshot of the full FOV of Doppler velocities estimated using the bisector method at an intensity level of $80\,$\%. Panel (b) shows the same snapshot, with filtered Doppler velocities, as described in \ref{sec:filteringandrebinning}. Black dashed circles highlight the excluded magnetic area with a sunspot, where all signals within the inner circle are masked. Panel (c) shows a Doppler-velocity time series at four different intensity levels taken from a single pixel, as indicated by a black `$\times$' in panel (b), for an arbitrary time period.}
    \label{fig:ibisdata}
\end{figure}

\begin{figure}[h!]
    \centering
    \plotone{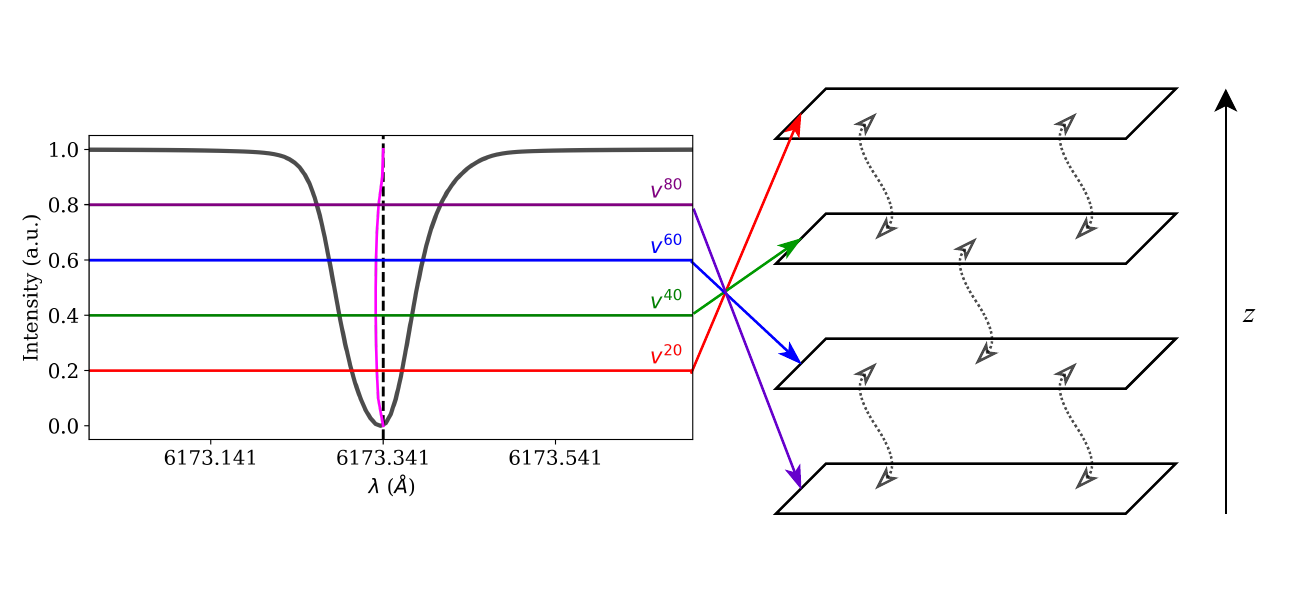}
    \caption{Illustration of estimating Doppler velocities using the bisector method. The left panel shows the absorption line profile of the $6173.34$\,\AA\ as a function of wavelength $I(\lambda)$. The line is then sampled at different intensity levels (relative to the continuum intensity, corresponding to $100\%$), $N$. The vertical line indicates the wavelength in the rest frame of reference (black, dashed) and the magenta line shows the calculated bisector of the line. The asymmetries caused by the Doppler shift at each level can then be translated to a Doppler velocity $v_{\mathrm{d}}^N$ at different heights $z$ within the solar atmosphere. This is illustrated on the right, highlighting how each intensity level corresponds to a different layer in the atmosphere. Curved lines illustrate propagating acoustic waves traveling between such layers. 
    }
    \label{fig:formationheightdrawing}
\end{figure}

Here, we adopt the same procedure described by \cite{2010ApJ...721L..86R} and \cite{ 2022ApJ...933..109Z} to estimate Doppler velocities as a function of intensity level $N$. First, the spectral line is interpolated and sampled at $10$ equally distributed intensities $N = I(\lambda)/\Ic$ with $N=(0, 10, ..., 90)\%$, relative to the continuum intensity $\Ic$. At each $N$, the line profile has two intensity values with respective wavelengths $\lambda_\mathrm{blue}$ and $\lambda_\mathrm{red}$. The shift from the rest frame of reference with $\lambda_0=6173.341\,$\AA\ at each $N$ is called the bisector. Doppler velocities $v_\mathrm{d}^N$ are calculated according to 
\begin{align}
    v_\mathrm{d}^N = \frac{\left( \lambda_\mathrm{blue} + \lambda_\mathrm{red} \right)/2 - \lambda_0}{\lambda_0}\cdot c\,,
\end{align}
where $c$ is the speed of light and $N=0$ corresponds to the line-core intensity level and therefore the largest height within the atmosphere. A snapshot of the IBIS observations, including an example of Doppler velocity time-series $v_\mathrm{d}^N$, is shown in Figure~\ref{fig:ibisdata} and similarly, results from the simulation data are shown in Figure~\ref{fig:simulationdata}. Note that for most visualizations in this study, we limit ourselves to displaying just four intensity levels and therefore the Doppler velocities $v_\mathrm{d}^{20}$, $v_\mathrm{d}^{40}$, $v_\mathrm{d}^{60}$ and $v_\mathrm{d}^{80}$. 
In general, Doppler velocities $v_\mathrm{d}^N$ are implicit functions of the atmospheric height $z$, thus, each $N$ has an approximate atmospheric layer corresponding to it, as illustrated in Figure~\ref{fig:formationheightdrawing}. 

\subsubsection{HMI-like algorithm}
Given the design of the MDI instrument onboard the {\it SOHO} observatory, estimating the Doppler velocity has to be achieved using spectral filter transmission functions \citep{1995SoPh..162..129S}. In the following, we will approximate the six different SDO/HMI \citep{2012SoPh..275..229S} filter profiles, which are similar to their MDI counterparts, and label this method HMI-like and the resulting Doppler velocity $v_\mathrm{like}^\mathrm{HMI}$. Following the procedure suggested by \citet{2012SoPh..278..217C}, we estimate the first Fourier components $a_1, b_1$ using the filtered line profile intensity as a function of wavelength $I(\lambda)$:
\begin{align}
    a_1 &= \frac{2}{T}\int_{-\frac{T}{2}}^{+\frac{T}{2}}I(\lambda)\cos\left( 2\pi\frac{\lambda}{T} \right)\mathrm{d}\lambda\\
    b_1 &= \frac{2}{T}\int_{-\frac{T}{2}}^{+\frac{T}{2}}I(\lambda)\sin\left( 2\pi\frac{\lambda}{T} \right)\mathrm{d}\lambda\,,
\end{align}
where $T=412.8\,$m\AA\ is the total wavelength domain. Now we compute $v_\mathrm{like}^\mathrm{HMI}$ according to 
\begin{align}
    v_\mathrm{like}^\mathrm{HMI} &= \frac{\mathrm{d}v}{\mathrm{d}\lambda}\frac{T}{2\pi}\mathrm{atan}\left( \frac{b_1}{a_1} \right)\,,
\end{align}
with $\mathrm{d}v/\mathrm{d}\lambda = 48562.4\,$m\,s$^{-1}$\,\AA$^{-1}$. 

As a last step, $v_\mathrm{like}^\mathrm{HMI}$ must be corrected using look-up tables, generated from a reference \ion{Fe}{1} line at rest (in this case, the reference line profile is obtained from a spatial average of all line profiles in the simulation domain shown in Figure~\ref{fig:formationheightdrawing}). By shifting the reference line profile in wavelength to simulate a Doppler velocity and convolving the profile for each shift with the HMI filter transmittance functions, the uncorrected $v_\mathrm{like}^\mathrm{HMI}$ is obtained. The look-up tables are therefore the inverse function of this process, such that the corrected $v_\mathrm{like}^\mathrm{HMI}$ can be obtained from interpolating the table.

\subsubsection{True velocity}
\begin{figure}[h!]
    \centering
    \includegraphics[width=7.1in]{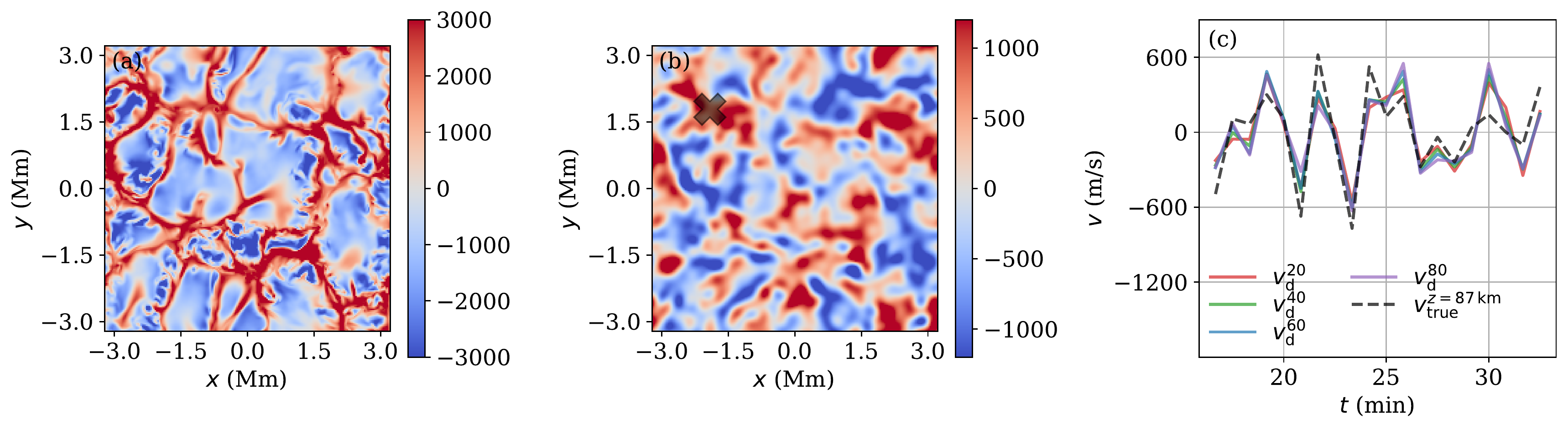}
    \caption{Vertical velocities extracted from the simulation dataset. Panel (a) shows a snapshot of the simulation box at a geometric height of $z=87$\,km. Panel (b) shows the corresponding filtered velocities similar to what is shown in Figure \ref{fig:ibisdata}. 
    Panel (c) shows a velocity time series, taken at the black $\times$ in panel (b) for four different intensity levels similar to Figure \ref{fig:ibisdata}. Additionally, the dashed black line shows the true velocity $v_\mathrm{true}^{z=87\mathrm{\,km}}$.
    }
    \label{fig:simulationdata}
\end{figure}

Aside from estimating Doppler velocities using the intensities that result from the line synthesis, we analyze the actual velocity field (or `true' velocities) from simulations. The oscillations are thus entailed in the true velocity vector $\mathbf{v}_\mathrm{true}$. Similar to the observational data, a snapshot plus an arbitrary time-series are shown in Figure \ref{fig:simulationdata}. The simulated vertical velocity shown in Figure \ref{fig:simulationdata}b is extracted at a geometrical height of $z=87\,$km, which roughly translates to an intensity level of $80\,$\%. It is not surprising to see both Doppler and true velocities behaving similarly in their variation trends, although differences, especially in amplitude, are also apparent.

In order to compare Doppler and true velocities, we consider the LOS component of $\mathbf{v}_\mathrm{true}$. Given the fact that $\alpha$ imitates a longitudinal location, considering the $z$ component $v_z$ as the vertical direction and the $y$ component $v_y$ as the horizontal direction is sufficient (see Figure \ref{fig:viewingangledrawing}). Because of the small horizontal size of the simulation box (6.4\,Mm), we neglect the latitudinal variation LOS and apply the same viewing angle to all pixels in the simulation domain. This simplifies the computation of the true LOS velocity to
\begin{align}
    v_\mathrm{true}^\mathrm{LOS}(\alpha) = v_z\sin\left(\alpha\right) - v_y\cos\left(\alpha\right)\,.
    \label{eq:v_los}
\end{align}

\subsection{Formation heights} \label{sec:formationheight}
\begin{figure}[h!]
    \centering
    \plotone{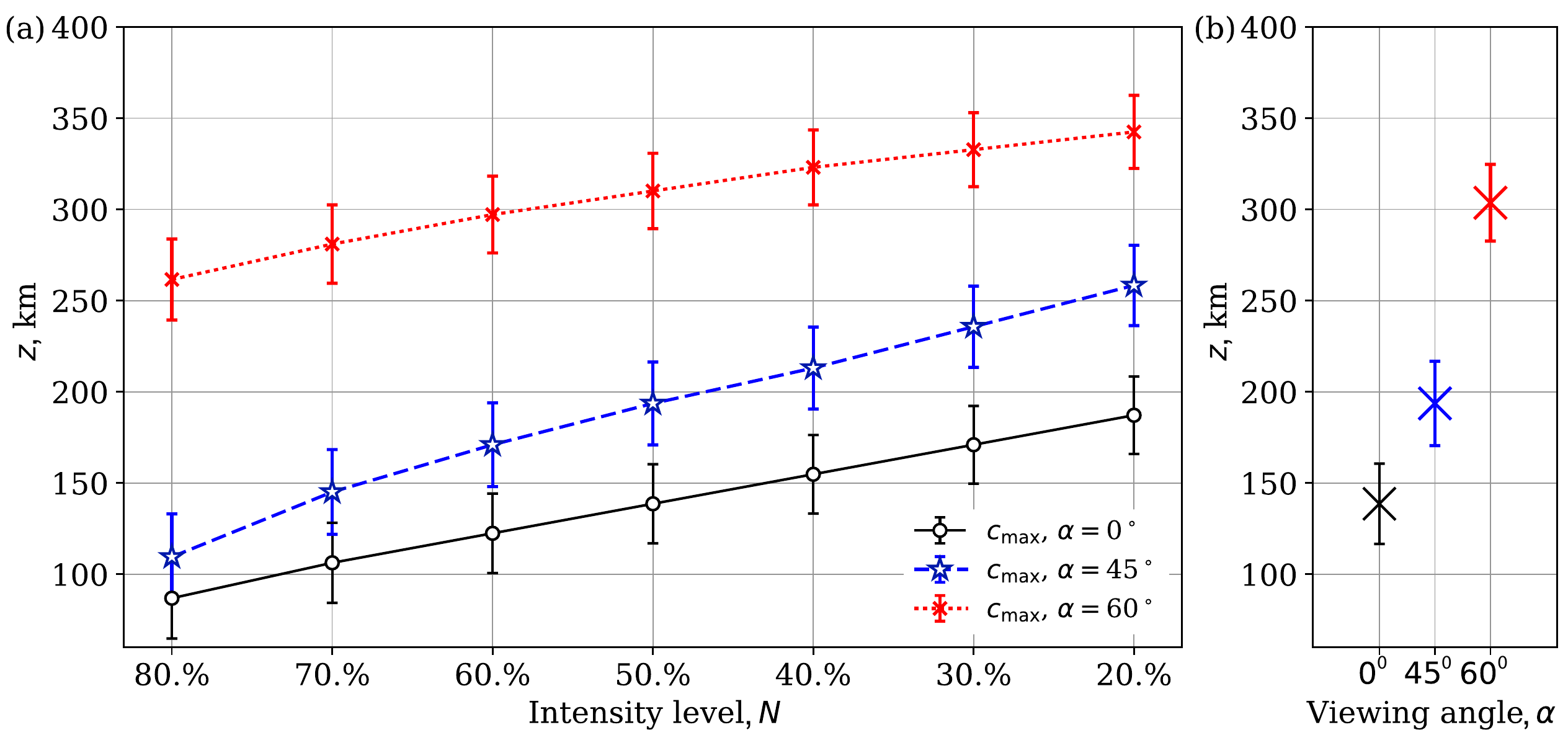}
    \caption{
    Estimations of the geometric height of the solar atmosphere obtained with two methods for three distances from the disk center: cross-correlation of the true velocity field with a) selected line profile levels and b) Doppler velocities obtained with the HMI algorithm.
    These relations allow us to obtain an approximate height $z$ for all the intensity levels used, called the equivalent height. Different colors indicate different viewing angles, $\alpha$.}
    \label{fig:formationheightall}
\end{figure}

Among different systematic effects influencing the helioseismic measurements mentioned earlier, variation of the line formation height plays an important, if not a dominant, role. When comparing Doppler velocities to true velocities, we have to ensure that they are extracted from the same atmospheric height. An established approach to this problem is demonstrated by \citet{2011SoPh..271...27F} and \citet{Kitiashvili2015}. The procedure can be described as selecting Doppler velocities $v_\mathrm{d}^N$ from one intensity level $N$ and finding its best match among all the true velocities $v_\mathrm{true}(z)$. Matching $v_\mathrm{d}^N$ for all $N$ thus gives us the intensity level as a function of height $N(z)$. Quantitatively evaluating the best match can be done by maximizing the Pearson correlation coefficient, $\mathrm{max}(c(\alpha, z))$, with
\begin{align}
    c(\alpha, z) = \frac{\displaystyle \sum_{x, y, t} \left[ \left( v_{\mathrm{d}}^N(\alpha) - \bar{v}_{\mathrm{d}}^N(\alpha) \right) \left( v_\mathrm{true}^\mathrm{LOS}(\alpha, z) - \bar{v}_\mathrm{true}^\mathrm{LOS}(\alpha, z) \right) \right] }
    {\sqrt{\displaystyle \sum_{x, y, t}(v_{\mathrm{d}}^N(\alpha) - \bar{v}_{\mathrm{d}}^N(\alpha))^2\sum_{x, y, t}(v_\mathrm{true}^\mathrm{LOS}(\alpha, z) - \bar{v}_\mathrm{true}^\mathrm{LOS}(\alpha, z))^2}}\,,
\end{align}
where $\sum\limits_{x, y, t}$ 
represents the sum over all pixels and time steps, and $\bar{v}$ is the mean of $v$ (averaged over $t$ in this case). Here we employ a cubic interpolation in $z$ to allow the formation height to fall between grid points of the $z$-axis. Finally, the maximum $c_\mathrm{max}(\alpha, z)=\mathrm{max}(c(\alpha, z))$ for both the bisector method and the HMI-like method are shown in Figure \ref{fig:formationheightall}.

The intensity-height relation shows the expected behavior, with the \ion{Fe}{1} 6173\,\AA\ forming between $65$--$208\,$km, the line core representing the highest layer in the atmosphere, larger viewing angles leading to a general shift towards larger line-forming heights and the HMI algorithm yielding an atmospheric height of roughly $116$--$160\,$km. These results also agree with the study by \citet{Kitiashvili2015}. Regarding error bars, we employ a Gaussian fit to $c(z)$ as a function of height and adopt the error of the correlation function $\sigma_c(z)$ to roughly equate the variance $\sigma_\mathrm{fit}(z)$ of the Gaussian. This procedure can only produce approximate results, since the true distribution of $c(z)$ is not Gaussian. Generally, we can assume that $\sigma_\mathrm{fit}(z)$ is slightly underestimated.

Although we show error bars for the intensity-height relation, it is important to understand that this does not mean Doppler velocities correspond to a single value height from which we can extract the true velocities. In fact, the error bars in Figure \ref{fig:formationheightall} can only represent how well $v_\mathrm{true}$ correlates to $v_\mathrm{d}$, not the actual height interval covered by the line formation process. Although the two may be related in some way, it must be kept in mind that the range in height that contributes to our given Doppler velocities cannot be known with certainty, unless we compute accurate (but nevertheless synthetic) intensity contribution functions. In the following we will represent $v_\mathrm{d}$ as forming at a single valued height, purely for notation purposes, but the uncertainty in formation height still applies to all Doppler velocities. 

\subsection{Filtering} \label{sec:filteringandrebinning}
In the recent study by \citet{2022ApJ...933..109Z}, a Gaussian-filter with a width of $\sigma=0\farcs637$ was used to smooth the data to enhance the signal-to-noise ratio. 
Similarly, we apply a Gaussian smoothing to the simulation data, albeit using a much narrower width of $\sigma=0\farcs022$ (1.28\,pixels), since the spatial scale of the simulation box is smaller and we have to ensure to capture the medium-sized oscillation modes.
Reducing the spatial resolution may strongly affect the estimation of the line-formation height \citep{Kitiashvili2015}. Here, we analyze different smoothing $\sigma$'s and conclude that changes in the formation height only start to become noticeable for $\sigma>0\farcs087$ ($\approx 5\,$pixel).

Given the fact that we are only interested in analyzing $p$-mode acoustic waves, a filter needs to be applied accordingly. This filter is designed to erase granulation and $f$-mode signals and is adopted from the data analysis procedures by \citet{2022ApJ...933..109Z} (see also Figure \ref{fig:power}). 

\subsection{Estimating power} \label{sec:estimatingpower}
\begin{figure}[h!]
    \centering
    \plotone{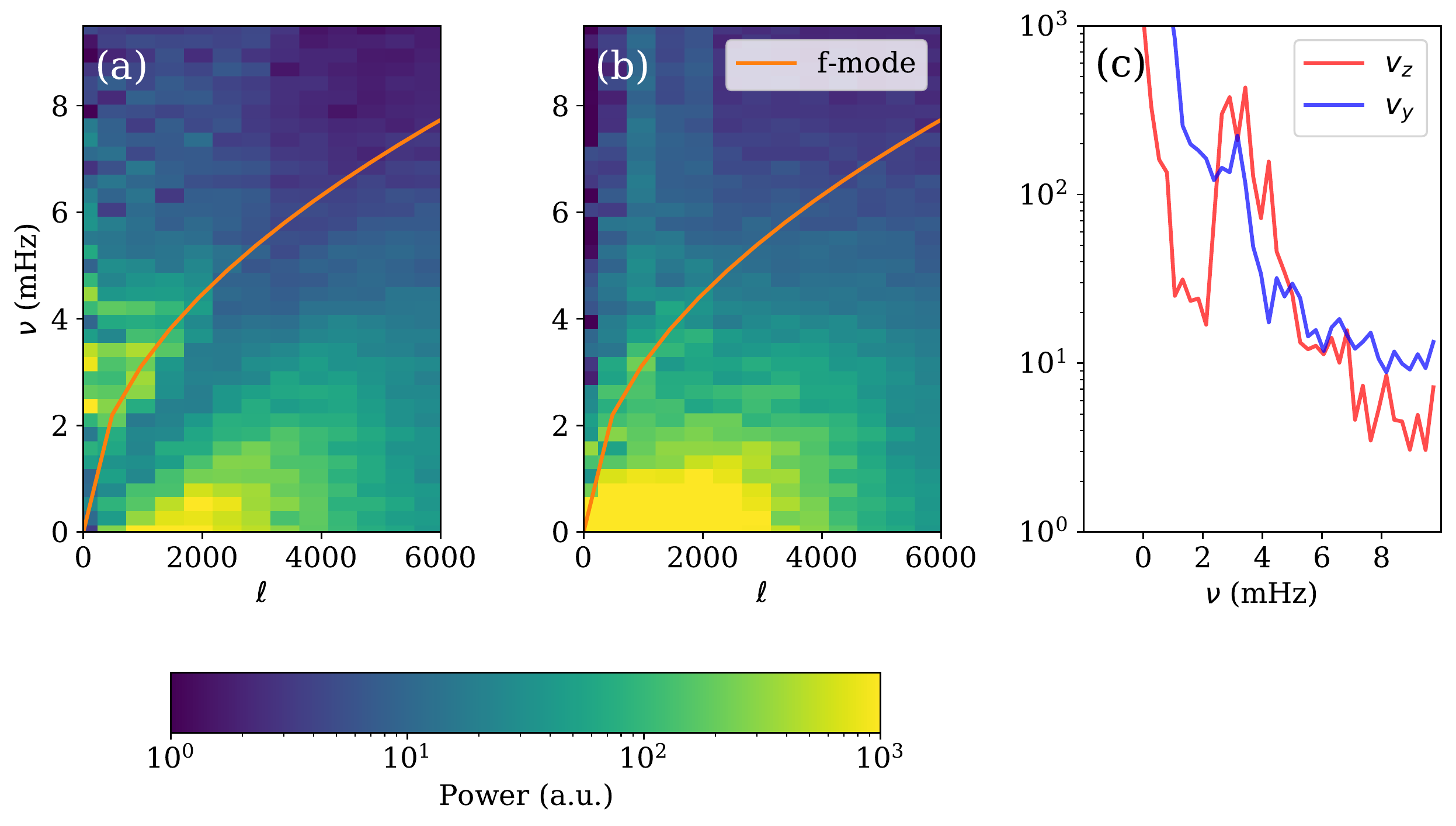}
    \caption{Power spectra $P(\ell, \nu)$ of vertical and velocities $v_z$ (a) and horizontal $v_y$ (b), both as functions of harmonic degree $\ell$ and temporal frequency $\nu$. The velocity fields are taken at a geometric height of $z=150\,$km. The $f$-mode ridge is shown as the orange curve in both plots. An arbitrary slice at $\ell = 965$ is shown in panel (c). 
    }
    \label{fig:power}
\end{figure}

Filtering is performed in the Fourier domain, for which we define $F\big(v(x, y, z, t)\big) = \widetilde{V}(k_x, k_y, z, \nu)$ as the Fourier-transform of $v(x, y, z, t)$, where $(k_x, k_y)$ are the horizontal wave numbers in $(x, y)$ direction. The spectral power $P$ is calculated as the absolute square of the $\widetilde{V}$:
\begin{align}
    P(\ell, \nu) = |\widetilde{V}(\ell, \nu)|^2,
\end{align}
where $\ell$ is the harmonic degree with $\ell = 2\pi R_\odot \sqrt{k_x^2 + k_y^2}$ with $R_\odot$ denoting the radius of the Sun. In Figure \ref{fig:power} we show the azimuthally-averaged power for both $v_z$ and $v_y$, highlighting the $f$-mode ridge. The $f$-mode ridge and all signals with lower frequencies are filtered out.

From the power spectra we see that the horizontal oscillation seen in $P(v_y)$ show large amplitudes even beyond the granulation domain. Generally we do not expect strong horizontal oscillation power in the acoustic domain. This is likely due to the excitation mechanism in the simulation and may be adjusted by modifying the compressibility. For viewing angles other than $\alpha=0^\circ$ it is therefore important to keep in mind that the horizontal component of the velocity has a large contribution to the LOS velocity (see Equation \ref{eq:v_los}).

\begin{figure}[h!]
    \centering
    \plotone{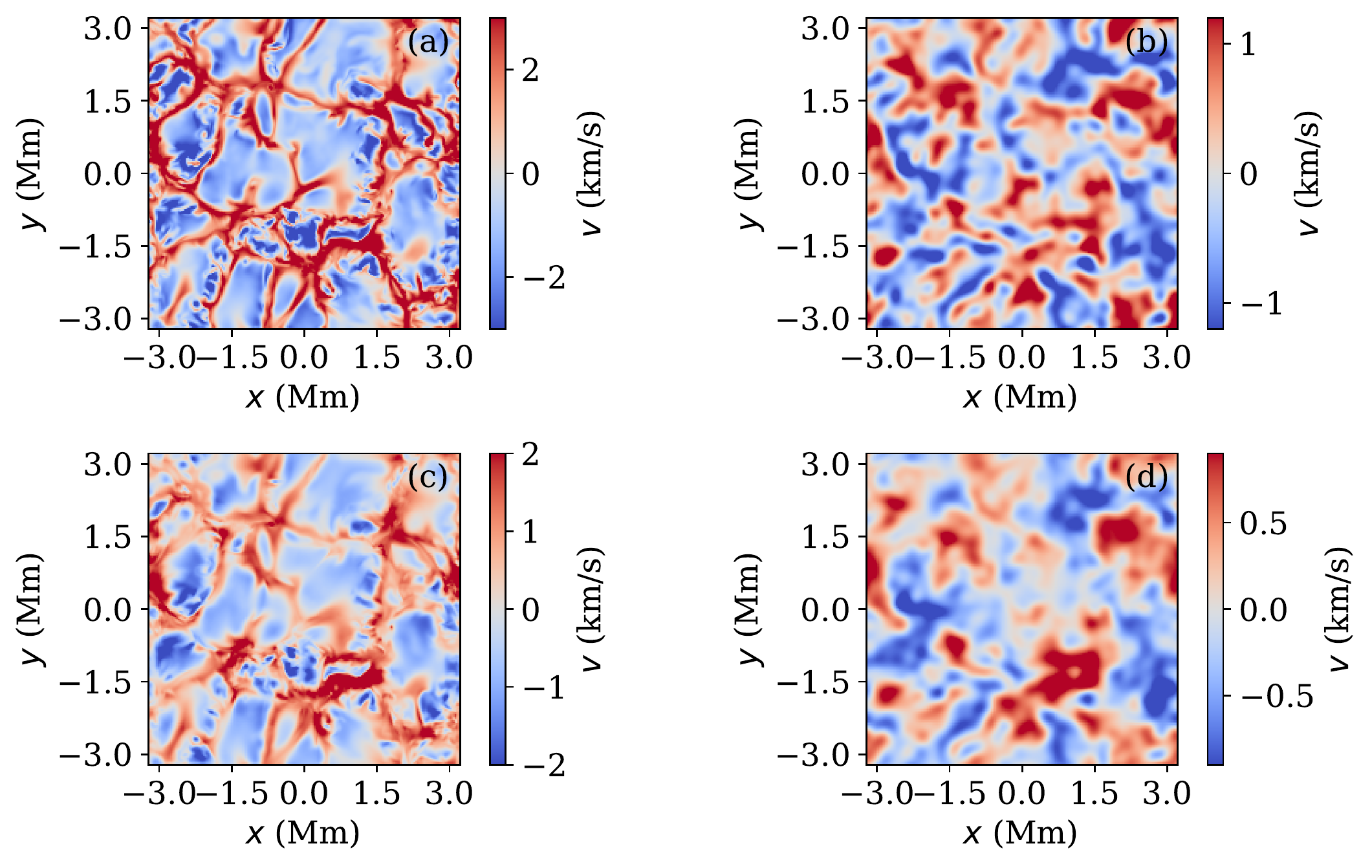}
    \caption{Doppler and true velocities from the simulation data, shown for equivalent heights and the same timestep. Panel (a) shows $v_\mathrm{true}^{z=87\mathrm{\,km}}$ before applying the f-mode filter, while panel (b) shows the same velocity field after the filter is applied. Panel (c) and (d) are the same but for $v_\mathrm{d}^{80}$.}
    \label{fig:simulationdata_afterfilter}
\end{figure}

The effects of the filter on the velocity fields is shown for one snapshot in Figure \ref{fig:simulationdata_afterfilter}. From this, it can be seen that small scale as well as long time period features, such as granulation and turbulent flows are removed.

\subsection{Estimating phase-shift} \label{sec:estimatingphases}
For this study, we compute the phase-differences $\delta\phi$ between different heights as functions of $(x, y)$ before performing a spatial average. After the velocities are filtered in the Fourier domain as mentioned above, they are spatially transformed back yielding $\widetilde{V}(x, y, z, \nu)$. From there, we compute phase-differences as
\begin{align}
    \delta\phi_{ij}(x, y, \nu) = \mathrm{arg}\left[ \widetilde{V}(x, y, z_i, \nu), \widetilde{V}(x, y, z_j, \nu) \right]\,.
    \label{eq:deltaphi}
\end{align}
Note that $z_i$ can also correspond to intensity levels $N$ according to Figure \ref{fig:formationheightall}. In order to reduce noise as much as possible, $\delta\phi_{ij}(x, y, \nu)$ is then spatially averaged. In the following we will use a more concise notation to denote phase-differences between different velocities. For example $\delta\phi[v_\mathrm{d}^{20}, v_\mathrm{true}^{z=87\,\mathrm{km}}]$ denotes $\delta\phi_{ij}(\nu)$ for $z_i \equiv N=20\%$ and $z_j = 87\,\mathrm{km}$. 

It must be noted that the spatial average of $\delta\phi_{ij}(x, y, \nu)$ is not always straightforward, since the $\mathrm{arg}$-function can result in values outside of the unit circle (i.e. $\delta\phi_{ij}(x, y, \nu) > \pi$ or $< -\pi$). Such occurrences therefore have to be treated in an extra step. This is done by discarding all the pixels $(x_0, y_0)$ where $\delta\phi_{ij}(x_0, y_0, \nu)$ is more than two standard-deviations $\sigma_{\delta\phi}$ away from the mean values, and the $\sigma_{\delta\phi}$ is derived from the $\delta\phi_{ij}(x, y, \nu)$ distribution for all pixels $(x, y)$ at constant $\nu$. After discarding the pixels showing outlier values in this way, the standard-deviation $\sigma_{\delta\phi}(\nu)$ of $\delta\phi_{ij}(x, y, \nu)$ is recomputed as a function of $\nu$, which then represents the error for $\delta\phi_{ij}(\nu)$.

\section{Results} \label{sec:results}
Although we expect that the phase-differences derived from the simulated velocities show differences compared to the observed  Doppler velocities, it is still of interest to gauge how different they compare to each other. A direct comparison is shown in Section \ref{sec:realdatavssimulateddata}.
When estimating phase shifts from observations, we are limited to using intensity measurements of line profiles for the computation of Doppler shifts, 
which means the resulting Doppler velocity is an implicit product of local temperature oscillations.
The true plasma motions, provided by the simulation data, may show systematic phase-shifts compared to Doppler oscillations as stated above.
We compare the phase-difference, $\delta\phi_{ij}(\nu)$, derived from Doppler velocities and from true velocities in Section \ref{sec:truevelocityvsdopplervelocity}. Finally, accurate helioseismic measurements rely on accurate estimates of systematic center-to-limb effects.
After this, we investigate the phase-differences in the lower atmosphere as a function of viewing angle $\alpha$, presented in Section \ref{sec:dependenceonviewingangle}.

\subsection{Observational data vs. simulated data} \label{sec:realdatavssimulateddata}
\begin{figure}[h!]
    \centering
    \includegraphics[width=7.1in]{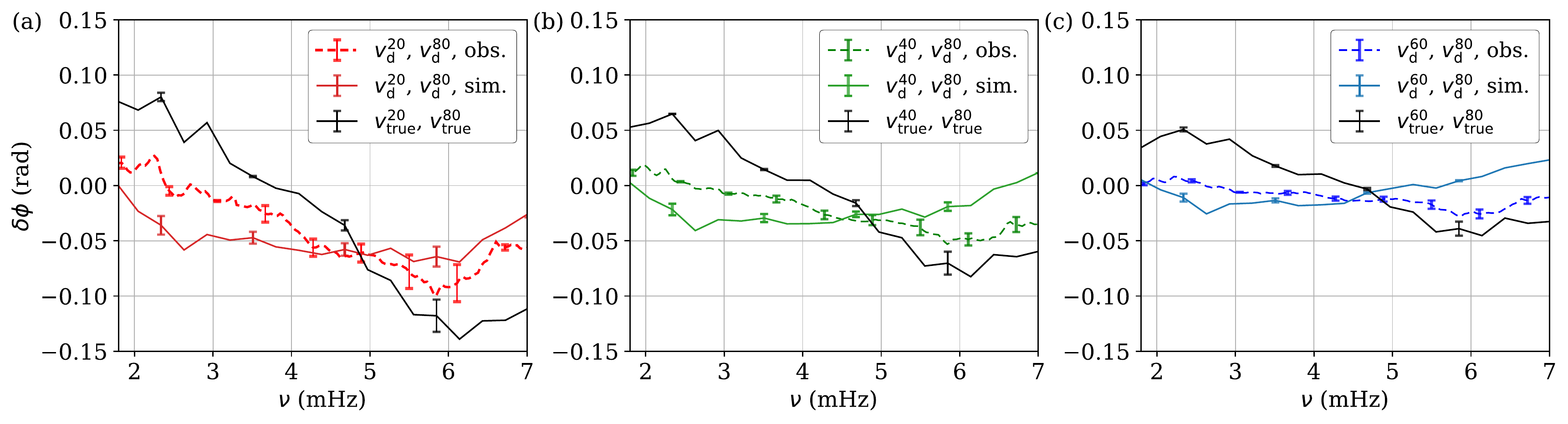}
    \caption{Phase-differences $\delta\phi$ at the disk center as function of frequency $\nu$ for Doppler velocities $v_\mathrm{d}$ derived from the bisector method for IBIS data (dashed curves), $v_\mathrm{d}$ derived from simulation data (solid), and true velocities $v_\mathrm{true}$ (black, solid curves). The phase-differences are shown for $\delta\phi[X]$ (from left to right): $X = [v_\mathrm{d}^{20}, v_\mathrm{d}^{80}]$, $[v_\mathrm{d}^{40}, v_\mathrm{d}^{80}]$, $[v_\mathrm{d}^{60}, v_\mathrm{d}^{80}]$.
    }
    \label{fig:phasedatavssim0deg}
\end{figure}

Comparing wave phases between observations and simulation data can best be done by looking at $\delta\phi_{ij}(\nu)$ derived from Doppler velocities $v_\mathrm{d}$, as well as true velocities $v_\mathrm{true}$ (for the simulation only). Here, we use $\delta\phi[X]$ with $X = [v_\mathrm{d}^{20}, v_\mathrm{d}^{80}]$, $[v_\mathrm{d}^{40}, v_\mathrm{d}^{80}]$, $[v_\mathrm{d}^{60}, v_\mathrm{d}^{80}]$. 
To extract the plasma velocities from simulations at the exact height values that match the equivalent intensity levels $N=20\%$, ..., $80\%$, we applied a cubic spline interpolation.
The comparison of the resulting phase-differences for simulated data and IBIS observations corresponding to the disk center is shown in Figure \ref{fig:phasedatavssim0deg}. 

Across all three panels in Figure \ref{fig:phasedatavssim0deg}, we find a decreasing trend in the overall amplitude of the $\delta\phi_{ij}(\nu)$ with decreasing height difference between the different layers (see also Figure \ref{fig:formationheightdrawing}).
This is expected in the sense that phase-differences between neighboring layers, at least for propagating waves, are smaller than phase-differences between far-apart layers. However the overall expectation is that waves are evanescent with little to no phase-differences. We can thereby confirm that the unexpected behavior reported by \citet{2022ApJ...933..109Z} is also seen in numerical simulations. Nevertheless, the results also show a clear disagreement between the simulated wave phases and the observed wave phases. 
While they are at similar levels in magnitudes, we see differences between $\delta\phi[v_\mathrm{d}, v_\mathrm{d}]$ and $\delta\phi[v_\mathrm{true}, v_\mathrm{true}]$ for all $\nu$. From one side, the phase difference computed from the simulated Doppler velocities especially shows uncharacteristic behavior, where $\delta\phi[v_\mathrm{d}^{20}, v_\mathrm{d}^{80}]$ is negative for all $\nu$ while both $\delta\phi[v_\mathrm{d}^{40}, v_\mathrm{d}^{80}]$ and $\delta\phi[v_\mathrm{d}^{60}, v_\mathrm{d}^{80}]$ show positive values for $\nu>6\,$mHz. Otherwise, $\delta\phi[v_\mathrm{true}, v_\mathrm{true}]$ qualitatively appears to be more in line with the observational data. The sign change occurs at around $4\,$mHz in true velocities, which is roughly $1\,$mHz larger compared to the observational data. In amplitudes, $\delta\phi[v_\mathrm{true}, v_\mathrm{true}]$ is roughly two to three times larger than $\delta\phi$ estimated from observations. 

For problems including dynamic processes in complex atmospheres, such as propagation of acoustic waves in the upper photosphere, it is extremely challenging to tune simulation parameters such that output data precisely match observations. Our main insight from Figure \ref{fig:phasedatavssim0deg} is that, qualitatively, both real and simulated data exhibit non-evanescent acoustic waves. 

\subsection{True velocity vs. Doppler velocity} \label{sec:truevelocityvsdopplervelocity}
\begin{figure}[h!]
    \centering
    \plotone{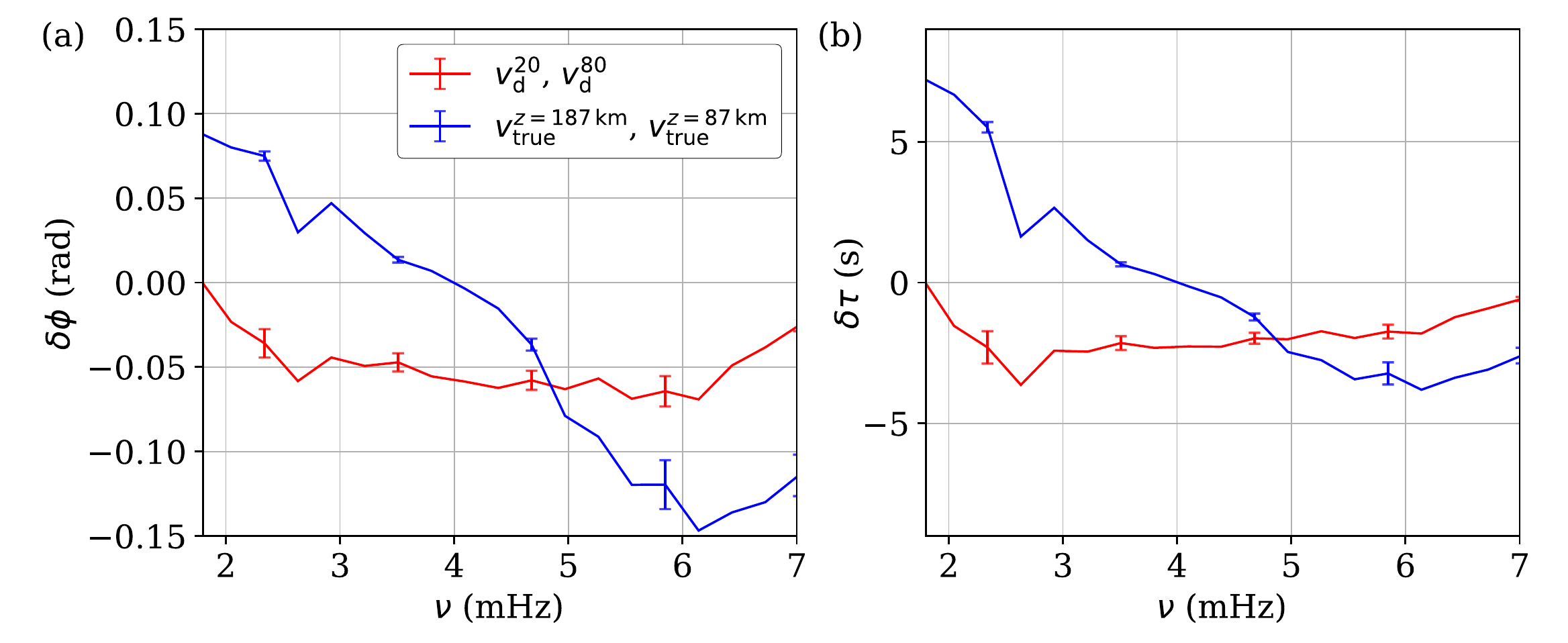}
    \caption{(a): Comparison between phase-differences $\delta\phi(\nu)$ for Doppler velocities (red) and true velocities (blue). Here, the vertical velocity $v_z$ is interpolated to match the approximate height of its corresponding intensity level $N$ (the equivalent height) according to Figure \ref{fig:formationheightall}.  (b) Travel-time difference $\delta\tau(\nu)$ corresponding to the $\delta\phi(\nu)$ in panel (a).}
    \label{fig:phasesoneheightcomparison}
\end{figure}

Discrepancies between true and Doppler velocities are expected in the lower solar atmosphere due to the aforementioned decoupling between perturbations in the gas and the according responses observed in intensity, and also due to differences in atmospheric heights (see Section \ref{sec:formationheight}). It is thus interesting to further investigate direct comparisons between $\delta\phi[v_\mathrm{d}, v_\mathrm{d}]$ and $\delta\phi[v_\mathrm{true}, v_\mathrm{true}]$ for the same height intervals. Here, this is done for $\delta\phi[v^{20}, v^{80}]$, in which case the equivalencies of $N = 20\%$ and $N = 80\% $ correspond to $z=187\,$km and $z=87\,$km heights, respectively, for the true velocities. The resulting phase-difference and travel-time difference, with $\delta\tau(\nu) = \delta\phi(\nu)/2\pi\nu$, are shown in Figure \ref{fig:phasesoneheightcomparison}.

Substantial difference between $\delta\phi[v_\mathrm{d}, v_\mathrm{d}]$ and $\delta\phi[v_\mathrm{true}, v_\mathrm{true}]$ is apparent in Figure \ref{fig:phasesoneheightcomparison}, confirming our aforementioned speculation. To restate our earlier point: in the lower solar atmosphere, the environment considered non-adiabatic can give rise to complex behaviors following perturbations in the atmosphere by, for example, an acoustic wave. It can thus be the case that such a perturbation is not instantaneously translated to an according modification in other properties of the atmosphere, such as temperature (and thus intensity). Therefore, the discrepancy observed between $\delta\phi[v_\mathrm{d}, v_\mathrm{d}]$ and $\delta\phi[v_\mathrm{true}, v_\mathrm{true}]$ could be an indicator to that such complex behaviors indeed affect Doppler velocities and their phases. Figure \ref{fig:phasesoneheightcomparison}a shows a qualitative agreement between phase-shifts for the Doppler shift and simulated velocities at different heights of the atmosphere. Quantitative discrepancies more than $6\,$s correspond to low frequencies. 

\begin{figure}[h!]
    \centering
    \plotone{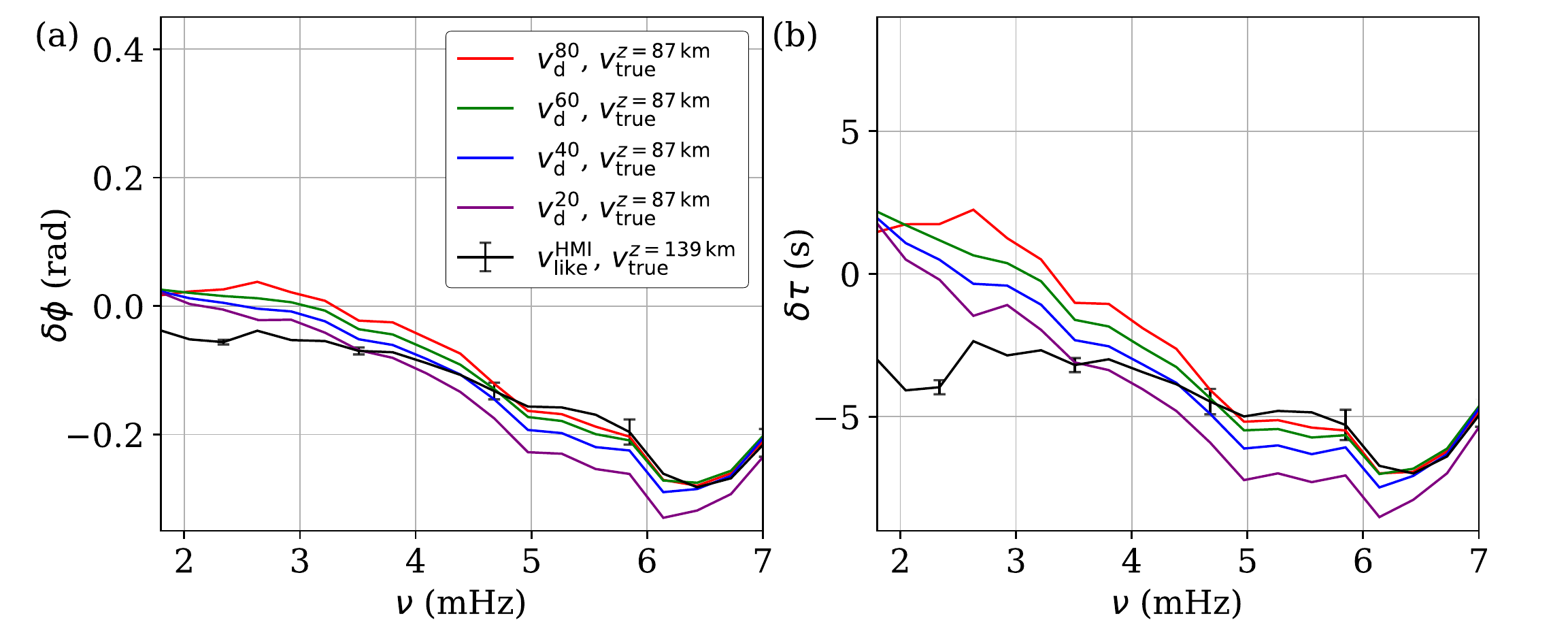}
    \caption{(a): Phase-differences $\delta\phi(\nu)$ obtained from Doppler velocities and true velocities for different heights. (b) Travel-time difference $\delta\tau(\nu)$ corresponding to $\delta\phi(\nu)$ shown in panel (a). For clarity, error bars are only shown for $\delta\phi[v_\mathrm{d}^\mathrm{HMI}, v_\mathrm{true}^{z=139\,\mathrm{km}}]$, and the errors are similar in other curves.}
    \label{fig:phasescross}
\end{figure}

Such a substantial discrepancy between $\delta\phi[v_\mathrm{d}, v_\mathrm{d}]$ and $\delta\phi[v_\mathrm{true}, v_\mathrm{true}]$ raises an important question: What does the phase-difference between $v_\mathrm{d}$ and $v_\mathrm{true}$, i.e. $\delta\phi[v_\mathrm{d}, v_\mathrm{true}]$, look like at the same atmospheric height? This phase-difference has to be treated carefully because the definitions of height are different for both quantities $v_\mathrm{d}$ and $v_\mathrm{true}$. Although we can consult our intensity-height relation, shown in Figure \ref{fig:formationheightall}, and extract $v_\mathrm{d}$, $v_\mathrm{true}$ at equivalent heights (again, interpolating $v_\mathrm{true}^{z}$), it is important to remember that $v_\mathrm{d}$ is formed in a large height-interval, while $v_\mathrm{true}$ can be extracted from a clearly defined $z$-value. With that in mind, we first compute $\delta\phi[v_\mathrm{like}^\mathrm{HMI}, v_\mathrm{true}^{z=139\,\mathrm{km}}]$ and $\delta\phi[v_\mathrm{d}^{80}, v_\mathrm{true}^{z=87\,\mathrm{km}}]$ to obtain two separate phase-differences at their equivalent heights. Additionally, $\delta\phi[v_\mathrm{d}^N, v_\mathrm{true}^{z=87\,\mathrm{km}}]$ at all other intensity levels are presented, such that the behaviors of $\delta\phi[v_\mathrm{d}, v_\mathrm{true}]$ can be studied for non-equivalent heights. The results are shown in Figure \ref{fig:phasescross}. 

An important observation from Figure \ref{fig:phasescross} is that the $\delta\phi[v_\mathrm{d}, v_\mathrm{true}]$ spectrum is seemingly dominated by a purely $\nu$-dependent term while variations from different heights are small, i.e. $\delta\phi[v_\mathrm{d}, v_\mathrm{true}](\nu)\approx f(\nu)$, where $f(\nu)$ does not depend on $z$ or $N$. This height-independent term $f(\nu)$ can be understood as another manifestation of the discrepancy between $\delta\phi[v_\mathrm{d}, v_\mathrm{d}]$ and $\delta\phi[v_\mathrm{true}, v_\mathrm{true}]$, indicating that the discrepancy is not only related to the differences in height definition, but also to other intrinsic differences between $v_\mathrm{d}$ and $v_\mathrm{true}$.

We can have a more intricate look at the wave propagation by computing phase-differences $\delta\phi(z)$ as a function of height. This is done by selecting a constant frequency $\nu_0$ (or a frequency band) and constructing $\delta\phi(z)$ as a list of $\delta\phi[v(z), v(z_0)]$ at constant frequency (or averaged over the selected frequency band).  Here we use $z_0 = 87\,$km$\equiv N=80\%$ and the frequency bands $\nu\in\left([1.8, 2.5], [3.5, 4.5], [4.5, 5.5], [6.5, 7.5]\right)\,$mHz. This gives $81$ data points for true velocities and $10$ data points for Doppler velocities. 

\begin{figure}[h!]
    \centering
        \includegraphics[width=7.in]{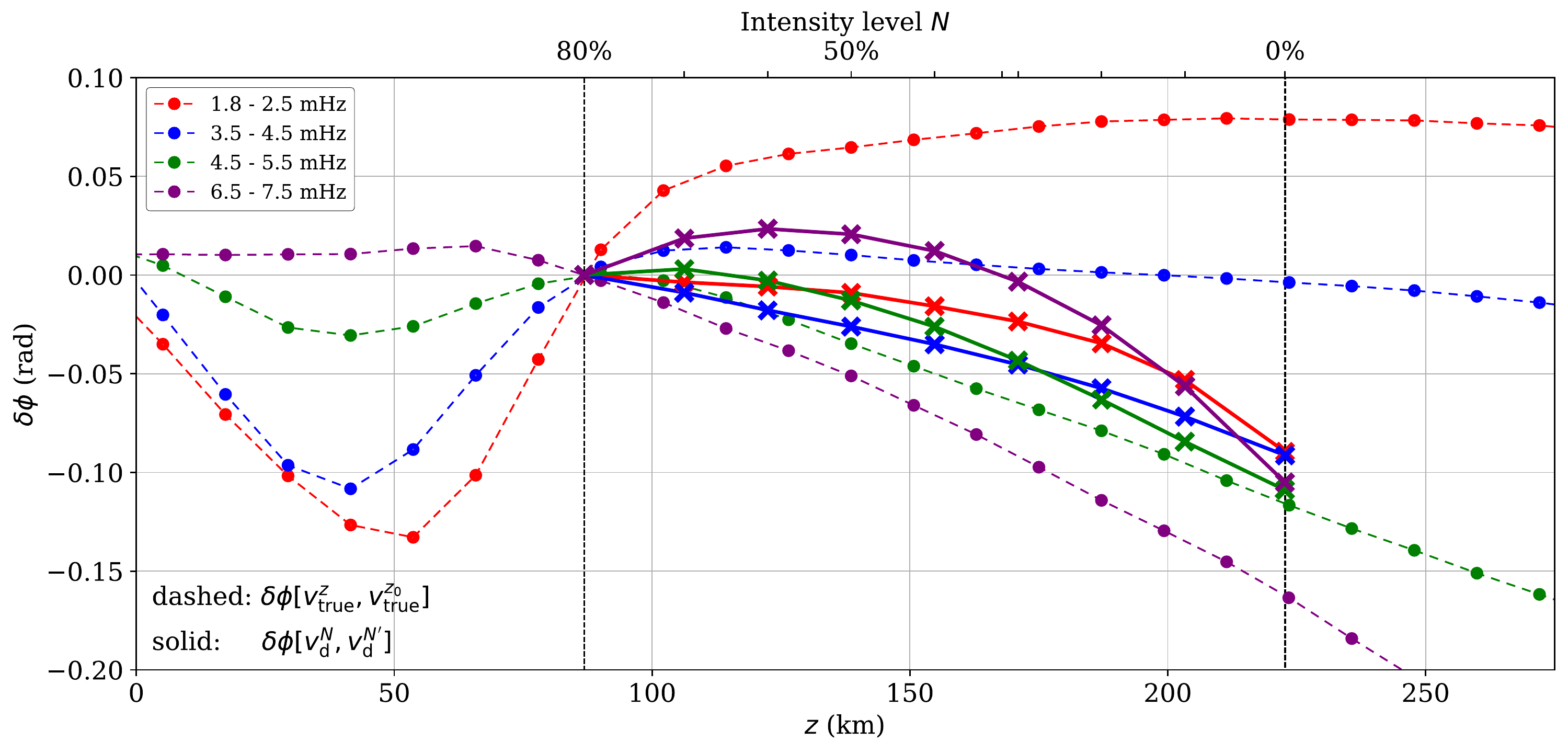}
    \caption{Comparison between phase-differences $\delta\phi$ for Doppler velocities (solid lines connecting crosses) and true velocities (dashed lines connecting dots). In this case, we show the growth of $\delta\phi(z)$ as functions of height ($N$ on the top axis and $z$ on the bottom axis). The height range spanned by $N\in[0\%, 80\%]$ is marked by the dark dashed lines. Each curve corresponds to an average over a frequency band as indicated in the legend. No error bars are shown here, as they are smaller than $1\%$.}
    \label{fig:crosspropagationvdopvsvzvzvzvdvd0deg}
\end{figure}

Analyzing the growth of phases as a function of height (Fig.~\ref{fig:crosspropagationvdopvsvzvzvzvdvd0deg}), our expectation for $\delta\phi[v^{z}, v^{z_0}]$ in the case of a fully adiabatic upper photosphere is a constant value of $0$ for $\nu\lesssim5.0\,$mHz and a monotonic decrease (if $z_0 < z$) for $\nu\gtrsim5.0\,$mHz. 
From Figure \ref{fig:crosspropagationvdopvsvzvzvzvdvd0deg}, we can see that these expectations are qualitatively met by $\delta\phi[v_\mathrm{true}^{z}, v_\mathrm{true}^{z_0}]$ for the frequency bands $\nu\in[3.5, 4.5]\,$mHz (blue), showing $\delta\phi\approx 0$ and $\nu\in[6.5, 7.5]\,$mHz (purple), showing said monotonic decrease towards larger heights.
Both of these frequency bands behave quite differently for $\delta\phi[v_{\mathrm{d}}^N, v_{\mathrm{d}}^{N^\prime}]$. Especially, the large $\nu\in[6.5, 7.5]\,$mHz band shows an odd trend, where an eventual decrease is proceeded by an initial growth. Similar to the results shown in Figure~\ref{fig:phasedatavssim0deg}, large as well as low frequencies seem to disagree the most between $\delta\phi[v_\mathrm{d}^N, v_\mathrm{d}^{N^\prime}]$ and $\delta\phi[v_\mathrm{true}(z), v_\mathrm{true}(z_0)]$. Nevertheless, almost all frequency bands show negative growth with increasing height toward decreasing negative $\delta\phi$ values (with the exception of $\nu\in[1.8, 2.5]\,$mHz for $\delta\phi[v_\mathrm{true}(z), v_\mathrm{true}(z_0)]$ ). This is a convincing observation of the non-evanescent nature of waves in the lower atmosphere, and the clear disagreement in phase shifts measured from Doppler and true velocities.

\subsection{Dependence on viewing angles} \label{sec:dependenceonviewingangle}

\begin{figure}[h!]
    \centering
    \plotone{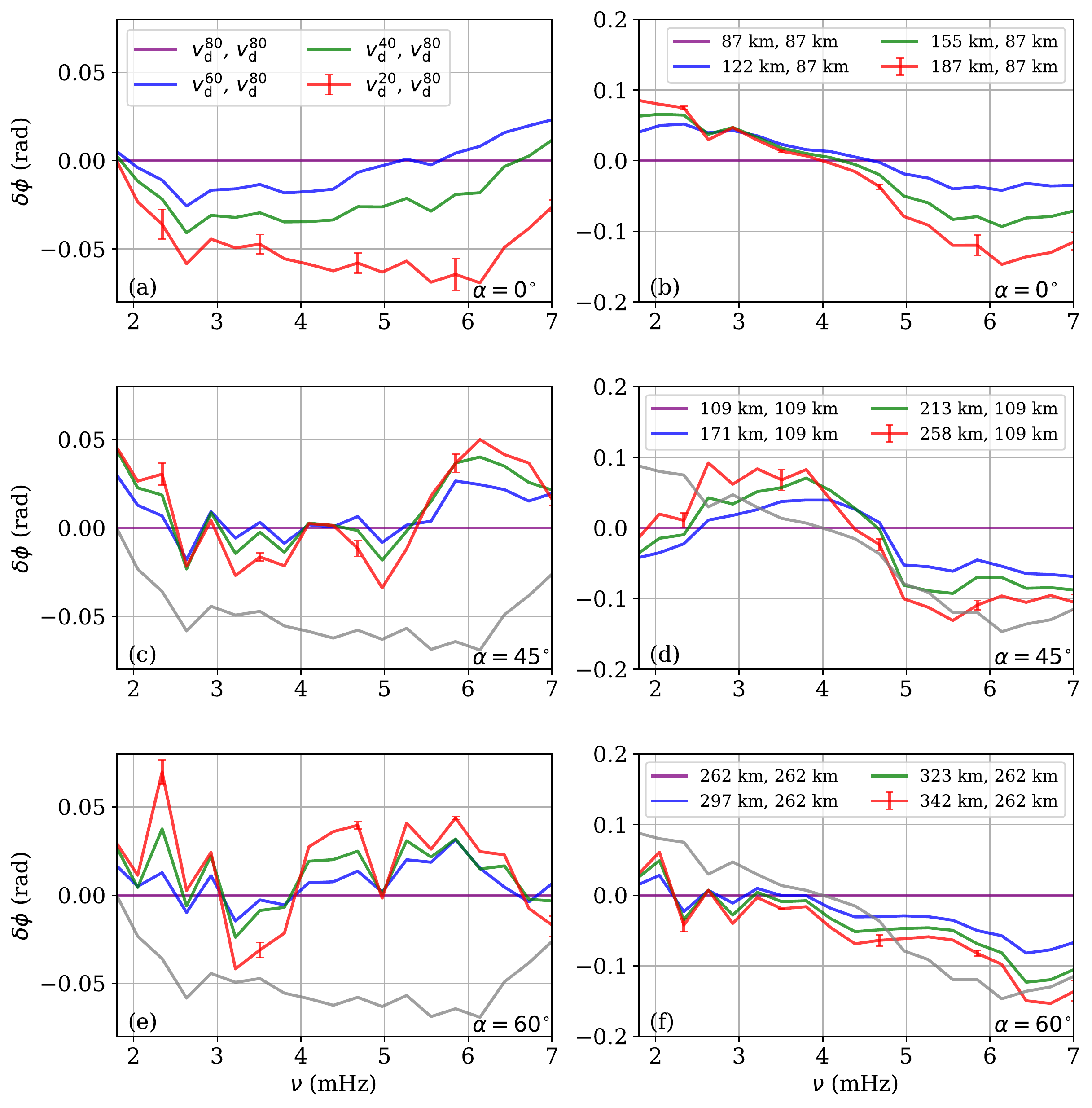}
    \caption{Phase-differences $\delta\phi(\nu)$ for both Doppler velocities ({\it left}) and true line-of-sight velocities ({\it right}) as functions of frequency $\nu$ for different viewing angles. The grey curve represents the phase-difference between the $20$\,\% and $80$\,\% layer at $\alpha=0^\circ$ for Doppler and true velocities respectively. For the true velocities, the base layers, to which the relative phase-differences are computed, are interpolated to match their corresponding intensity levels. For this Figure, only the respective atmospheric heights are displayed in the legend.}
    \label{fig:phasesvdvsvzalldegs}
\end{figure}

Frequency-dependent systematic over- or under-estimates of helioseismic travel times due to the center-to-limb variations have been analyzed by \citet{2018ApJ...853..161C}. Such systematic effects stem presumably from oscillation signals being measured at various heights, as travel-times are measured across the solar disk. It it thus expected that variations in viewing angles (and therefore heights) have an effect on the phase shifts of acoustic wave in the lower atmosphere as well. To further investigate this, we compute phase-differences as functions of viewing angle $\delta\phi(\alpha)$ for both Doppler and true velocities (Fig.~\ref{fig:phasesvdvsvzalldegs}). 

A first observation that we can make from Figure \ref{fig:phasesvdvsvzalldegs} is that indeed, $\delta\phi$ does strongly depend on $\alpha$. Hereby, $\delta\phi[v_\mathrm{d}^N, v_\mathrm{d}^{N^\prime}]$ shows roughly similar amplitudes across all $\alpha$, although it varies erratically around $0$ for $\alpha=45^\circ$ and $60^\circ$. For the true velocities, in the case of $\alpha=45^\circ$ we observe a qualitative increase in the frequency at which the sign change occurs, and for $\alpha=60^\circ$ a weaker amplitude at lower frequencies can be seen. A trend toward lower sign-reversal frequencies reported by \citet{2018ApJ...853..161C} cannot be observed here, although the signal is too erratic to confirm this behavior.

\begin{figure}[h!]
    \centering
    \includegraphics[width=7.in]{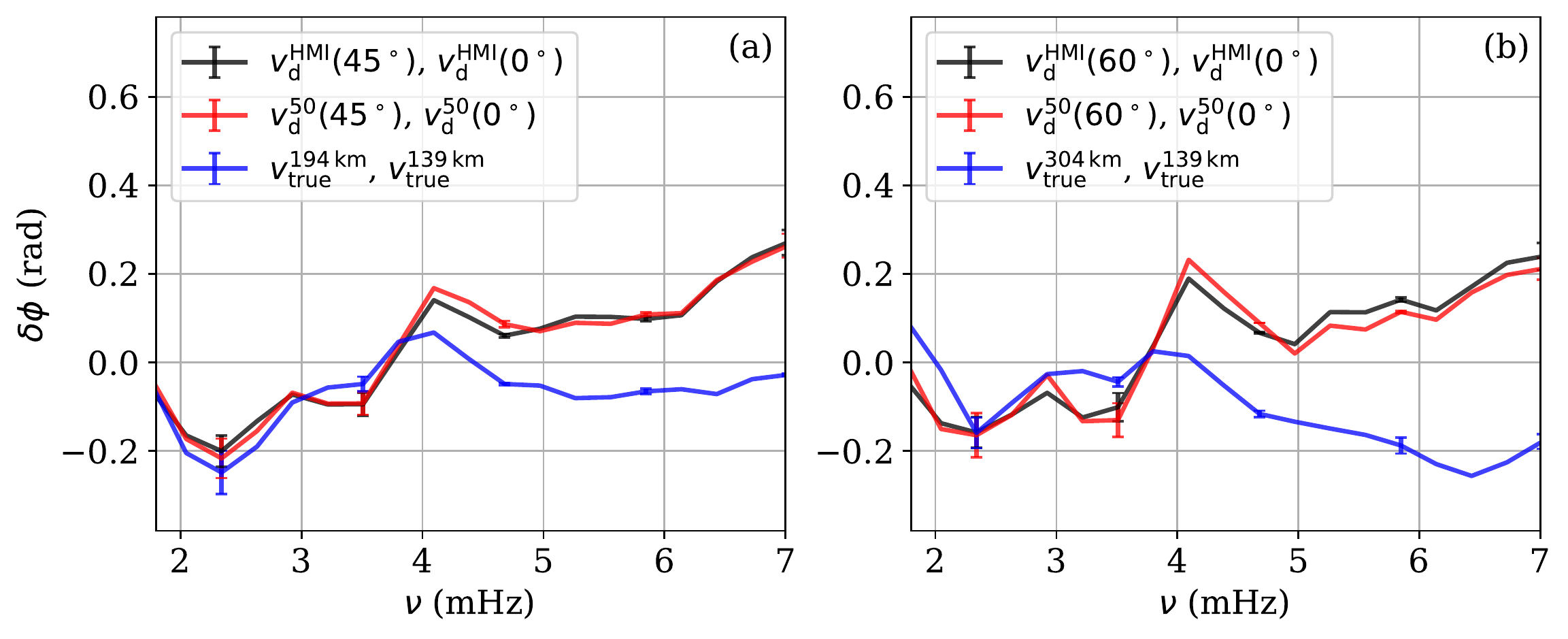}
    \caption{Phase-difference $\delta\phi(\nu)$ between viewing angles $\alpha=45^\circ$ (a), $\alpha=60^\circ$ (b) with  $\alpha=0^\circ$, as functions of frequency $\nu$. These are shown for Doppler velocities (both from the bisector (red) and the HMI-like (black) estimation methods) and true line-of-sight velocities (blue). Doppler velocities derived via the bisector method are extracted at intensity level $N=50\%$ to match the HMI line formation height as closely as possible (see also Figure \ref{fig:formationheightall}). Note that the true velocities are interpolated to match their heights to the corresponding intensity levels and the formation heights of the Doppler velocities.  
    }    
    \label{fig:phasesXdegvs0deg}
\end{figure}

Finally, it is important for helioseismic measurements to estimate the spurious phase-shifts between two spatially separated points. Here we imitate two separate disk locations by computing $\delta\phi[v^{\alpha}, v^{\alpha=0^\circ}]$, which represents the phase-difference between a region close to disk center and a region close to the limb viewed at an angle of $\alpha$ (see Figure \ref{fig:viewingangledrawing}). The result for $\alpha=45^\circ, 60^\circ$ is shown in Figure \ref{fig:phasesXdegvs0deg}.

As expected, both $\delta\phi[v_\mathrm{d}, v_\mathrm{d}]$ from the bisector and the HMI-like method behave very similarly. Interestingly, both curves show a slight deviation from the behavior that is observed in the analysis so far: If we assume that the non-zero phase-difference observed for, e.g., $\delta\phi[v_\mathrm{d}^{\alpha=45^\circ}, v_\mathrm{d}^{\alpha=0^\circ}]$ is solely due to the height difference in the respective atmosphere, we could argue that the frequency-dependent trend should roughly show the same behavior as the green or blue curve in Figure \ref{fig:phasesvdvsvzalldegs} (panel a). However, we find negative values only for $\nu<4\,$mHz followed by an increase toward positive values for larger $\nu$.
Furthermore, the difference between $\delta\phi[v_\mathrm{d}^{\alpha=45^\circ}, v_\mathrm{d}^{\alpha=0^\circ}]$ and $\delta\phi[v_\mathrm{d}^{\alpha=60^\circ}, v_\mathrm{d}^{\alpha=0^\circ}]$ should follow the same trend that is observed when moving from green to blue to purple (in Figure \ref{fig:phasesvdvsvzalldegs}a), i.e. larger negative phase-differences and no positive valued phase-differences. This is not observed for Figure \ref{fig:phasesXdegvs0deg}. Instead, we find roughly the same values for most $\nu$. Both of these facts indicate that the difference in atmospheric height is not the only cause of spurious phase-shifts. Likely it is also the case that horizontal velocity components contribute to the differences, as they are larger than expected (see Figure \ref{fig:power}).

Finally, $\delta\phi[v_\mathrm{true}, v_\mathrm{true}]$ surprisingly agrees with $\delta\phi[v_\mathrm{d}, v_\mathrm{d}]$ at lower frequencies, but exhibits no significant positive values at larger frequencies. This is also not necessarily what is expected, based on results so far, at least at lower frequencies, for the same reasons stated above.

\section{Discussion} \label{sec:discussion}
In this study, we have used three-dimensional realistic and fully radiative simulations to investigate phase-shifts of acoustic waves in the lower photosphere. Simulated data provide true motions in the atmosphere and thus offer an opportunity of investigating the nature of the unexpected phase-shifts reported by \citet{2022ApJ...933..109Z}. A direct comparison between $\delta\phi$ derived from simulated data and observations can serve to validate the accuracy of our simulated data. However, let us reiterate that simulations and observations behaving differently is not surprising in general. Simulations can only yield a picture with limited realism, given the fact that we cannot capture all the physics present in the real Sun. Observations on the other hand cannot be inherently wrong, however, systematic effects due to instrumental influences or an incomplete understanding of measurement procedures can always tamper with the final data. At any rate, these results still present an advancement in understanding effects due to varying line-formation heights and non-evanescence in the lower atmosphere. The results of this work can roughly be divided into three parts, each containing a point of major insight and summarized in the following.

\subsection{Non-evanescent behavior in simulated data}
Acoustic waves clearly show non-evanescent behavior below the presumed cutoff frequency in the simulated data presented here. Figure \ref{fig:phasedatavssim0deg} highlights the fact that both $\delta\phi[v_\mathrm{d}, v_\mathrm{d}]$ and $\delta\phi[v_\mathrm{true}, v_\mathrm{true}]$ deviate from 0 for most frequencies and therefore confirms the non-evanescent nature of acoustic waves present in the simulated photosphere. Still, there are clear discrepancies between phase-differences derived from simulations and observations, which merit speculations. Looking at Figures~\ref{fig:ibisdata} and \ref{fig:simulationdata}, we can see that, qualitatively, the velocities behave very similarly. Derivations of the simulated velocities are based on synthesizing the \ion{Fe}{1} absorption line, which is a challenging task. Although \citet{Kitiashvili2015} argued a good agreement between the simulation and observational data, inherent differences between the two datasets are unavoidable. 

Moreover, the spatial sampling used in our simulation data, for example, is far smaller than what is available in observations. By implication, the actual oscillation modes that contribute to the estimation of $\delta\phi$ (see Equation \ref{eq:deltaphi}) can be quite different between the two datasets. In the case of our simulation, very large $\ell$ dominate the power spectrum (in Figure \ref{fig:power}, left panel) and oscillations with $\ell<500$, which contains most of solar oscillation power, are barely resolved. Other reasons include the presence of magnetic fields, which can not be fully removed from observational data. Finally, as mentioned, identifying formation heights, i.e. estimating the intensity-height relation shown in Figure \ref{fig:formationheightall} is a challenging task. An exact accordance to real data can hardly be achieved. Knowing how sensitive calculations of phase-differences can be against atmospheric heights and how spectral lines are formed over a range of different heights, we have to keep in mind that direct comparisons as shown in Figure \ref{fig:phasedatavssim0deg} may be inaccurate at best and misleading at worst.

\subsection{Discrepancies between phases from Doppler velocities and true velocities}
Another interesting insight can be gained from Figure \ref{fig:phasesoneheightcomparison}, which shows phase-differences between the Doppler and true velocities. In order to make the phase-differences as comparable as possible, we interpolated the true velocities to match the heights where the corresponding intensities are formed. Evidently, $\delta\phi[v_\mathrm{true}, v_\mathrm{true}]$ and $\delta\phi[v_\mathrm{d}, v_\mathrm{d}]$ are substantially different from each other. 
It is important to keep in mind that the computation of phase-shifts between different layers may show an implicit difference between geometrical height $z$ and intensity levels $N$. Notably, the phase-shifts obtained between different geometrical heights of the solar atmosphere capture dynamics from a wider range of optical depths. Therefore, the variation of turbulence strength between the geometrical height in intergranular lanes and the optical depth, which corresponds to lower atmospheric height in comparison with granulation, likely introduces a difference between $\delta\phi[v_\mathrm{true}, v_\mathrm{true}]$ and $\delta\phi[v_\mathrm{d}, v_\mathrm{d}]$. The effect may be amplified due to much stronger flows in the intergranular lanes and vortical downflows.
Nevertheless, both $\delta\phi[v_\mathrm{true}, v_\mathrm{true}]$ and $\delta\phi[v_\mathrm{d}, v_\mathrm{d}]$ are in qualitative agreement with the observational data in its variation trend with frequency (Figure \ref{fig:phasedatavssim0deg}), although it is noteworthy that the sign reversal occurs at roughly $4\,$mHz in the simulation data while it occurs near $3\,$mHz in the observation data. 

Results shown in Figure \ref{fig:phasescross} indicate that phase-differences between the velocities themselves, $\delta\phi[v_\mathrm{d}, v_\mathrm{true}]$, are dominated by a frequency-dependent factor $f(\nu)$ whereby variations in height contribute significantly less to the $\delta\phi[v_\mathrm{d}, v_\mathrm{true}]$ than $f(\nu)$. A similar result would be expected for phase-differences with a large height separation causing a similar spectrum to $f(\nu)$ and then introducing only small variations in height, which would explain the rest of the observed spectrum $\delta\phi[v_\mathrm{d}, v_\mathrm{true}](\nu) - f(\nu)$. In other words, a large base-discrepancy between $v_\mathrm{d}$ and $v_\mathrm{true}$ causes most of the contributions to the resulting phase-difference $\delta\phi[v_\mathrm{d}, v_\mathrm{true}]$, independent of different $N$. 

Following the hypothesis given in \citet{2022ApJ...933..109Z} (as summarized in Section \ref{sec:introduction}) we indeed expect a phase-shift between $v_\mathrm{d}$ and $v_\mathrm{true}$, but we also expect little to no phase-shift between $v_\mathrm{true}$ of different heights (for $\nu<5\,$mHz). The simulation nonetheless exhibits non-zero $\delta\phi[v_\mathrm{true}, v_\mathrm{true}]$, although we neither know whether this is expected in observation data (where $\delta\phi[v_\mathrm{true}, v_\mathrm{true}]$ corresponds to the phase of the wave-introduced gas perturbations), nor what causes $\delta\phi[v_\mathrm{true}, v_\mathrm{true}]\ne 0$ for the simulation data. It is worth mentioning that internal gravity waves can carry wave energy upward and yield a positive contribution to the phase-difference at low frequencies \citep[c.f.][]{rsta.2020.0170}. However, it is unclear if such an interpretation applies to waves in the evanescent part of the spectrum.

\subsection{Dependence of $\delta\phi$ on the viewing angle $\alpha$}
Finally, we explore how different viewing angles affect our phase-difference estimation. In principle, moving toward larger viewing angles, which can be translated to moving toward the solar limb (see Figure \ref{fig:viewingangledrawing}), corresponds to a shift in atmospheric heights. This shift in height is most likely the dominating cause to the helioseismic center-to-limb effect \citep{2018ApJ...853..161C}. We can translate different viewing angles to atmospheric heights using the intensity-height relation (Figure \ref{fig:formationheightall}), for example, Doppler velocities computed at $N=30\,\%$ and $\alpha=0^\circ$ corresponds roughly to the same geometric height $z$ as $N=60\,\%$ and $\alpha=45^\circ$. Now, in our simulations, we find that this one-to-one translation between $\alpha$ and $z$ is too simplistic to predict the cross-propagation (between different viewing angles) shown in Figure \ref{fig:phasesXdegvs0deg}: Looking at the red curve in panel (a), representing $\delta\phi[v_\mathrm{d}^{20}(\alpha=45^\circ), v_\mathrm{d}^{80}(\alpha=0^\circ)]$, we find a substantially different behavior compared to the blue curve in panel (c) of Figure \ref{fig:phasesvdvsvzalldegs}, representing $\delta\phi[v_\mathrm{d}^{60}(\alpha=45^\circ), v_\mathrm{d}^{80}(\alpha=45^\circ)]$. Here the equivalent height $z$ does not vary much ($87\,$km to $187$\,km for the red curve vs. $109\,$km to $171$\,km for the blue curve). By implication, the variation in $\alpha$ must be the explanation to the different behaviors. 

A possible explanation can be posed by the unexpectedly large horizontal oscillatory power (see Figure \ref{fig:power}, center and right panel), yielding significant contributions to the estimates of $\delta\phi[v_\mathrm{d}^{\alpha=45^\circ}, v_\mathrm{d}^{\alpha=0^\circ}]$ and $\delta\phi[v_\mathrm{d}^{\alpha=60^\circ}, v_\mathrm{d}^{\alpha=0^\circ}]$. Another factor to consider is the intensity contribution function likely strongly varying with $\alpha$: At $\alpha=0^\circ$ the spectral line is made up of intensity contributions from a height range of roughly $[150, 350]\,$km \citep{2011SoPh..271...27F}. While the contribution function for $\alpha=45^\circ$ will be shifted toward larger $z$, the mentioned height interval also increases. This can be seen in $z(N, \alpha=60^\circ)$ (Figure \ref{fig:formationheightall}) being much flatter than $z(N, \alpha=0^\circ)$. Thus, Doppler velocities at larger $\alpha$ correspond to increasingly larger height intervals, which may introduce additional phase-shifts as observed in Figure \ref{fig:phasesXdegvs0deg}. 
Another consideration is the emission integrated along the LOS, which will have contributions expanding into horizontal directions as $\alpha$ increases. Any effect from this is expected to be small however, since the relevant vertical height is below $200\,$km, therefore limiting the horizontal direction to just a few pixels and such small scale effects are removed by the acoustic filter.
Lastly, other effects may cause the $\alpha$-dependent phase-shifts, such as the convective blueshift \citep{2012ApJ...760L...1B}, which describes observations of Doppler velocities being increasingly dominated by granular upflows (as opposed to intergranular downflows) as we move to larger $\alpha$.

\section{Conclusion} \label{sec:conclusion}
As recent observations show progress in advancing our understanding of the helioseismic center-to-limb effect, we utilize state-of-art simulations to study the phase behavior of acoustic waves in the lower solar atmosphere. Roughly dividing this work into three parts, we have compared phase-differences $\delta\phi$ measured from IBIS observations and from \textit{StellarBox} simulations, studied the different behaviors of phases derived from Doppler and true velocities a 3D numerical model, and investigated the dependence of phase-differences $\delta\phi$ on the viewing angle $\alpha$. In summary, our three major insights are:
\begin{itemize}
    \item [1.] Acoustic waves show non-evanescent behaviors in the simulation presented here. 
    Further investigation of this phenomenon needs spectrally-resolved oscillation measurements at different disk locations. Suitable instruments for such observations include BBSO/GST \citep{2010AN....331..620G} as well as HELLRIDE \citep{2020SPIE11447E..AMP}.
    \item [2.] Between Doppler and true velocities, $\delta\phi[v_\mathrm{d}, v_\mathrm{d}]$ and $\delta\phi[v_\mathrm{true}, v_\mathrm{true}]$ show substantial differences relative to each other,
    as was speculated by \citet{2022ApJ...933..109Z}. However, we can not answer why $\delta\phi[v_\mathrm{true}, v_\mathrm{true}]$ is non-zero within the scope of this work. Evidently, more simulated scenarios, ideally using different simulation codes,
    are needed to ensure that simulation artifacts do not play a role in this result.
    \item [3.] Phase-differences $\delta\phi$ show a strong dependence not only on $\nu$, but also on $\alpha$, i.e. disk location.
\end{itemize}

Ultimately, studies of this nature allow us to further understand and quantify systematic effects affecting helioseismic measurements and therefore will allow us to determine the deep meridional-flow profile more precisely.\\
\\
{\bf Acknowledgments.} This work is partly supported by a NASA Heliophysics Supporting Research grant 80NSSC19K0857. At the time the IBIS observation was acquired, Dunn Solar Telescope at Sacramento Peak, New Mexico was operated by National Solar Observatory (NSO). NSO is operated by the Association of Universities for Research in Astronomy (AURA) Inc.~under a cooperative agreement with the National Science Foundation. 


\bibliography{lit.bib}{}
\bibliographystyle{aasjournal}

\end{document}